\documentclass{pasj00}
\begin{document}
\SetRunningHead{K.Ando et al.}{Astrometry of ON2N with VERA}
\Received{2010/07/30}%{yyyy/mm/dd}
\Accepted{2010/12/24}%{yyyy/mm/dd}
%\Published{}%{yyyy/mm/dd}

\title{Astrometry of Galactic Star-Forming Region ON2N with VERA:
       Estimation of the Galactic Constants}

%%% begin:list of authors
% Do NOT capitalize all letters in "textsc".
\author{Kazuma \textsc{Ando},\altaffilmark{1}
Takumi \textsc{Nagayama},\altaffilmark{2}
Toshihiro \textsc{Omodaka},\altaffilmark{1}
Toshihiro \textsc{Handa},\altaffilmark{3} 
\thanks{Present address: Graduate School of Science and Engineering, Kagoshima University,
        1-21-35 K\^orimoto, Kagoshima, Kagoshima 890-0065}\\
Hiroshi \textsc{Imai},\altaffilmark{1}
Akiharu \textsc{Nakagawa},\altaffilmark{1}
Hiroyuki \textsc{Nakanishi},\altaffilmark{1}\\
Mareki \textsc{Honma},\altaffilmark{2}
Hideyuki \textsc{Kobayashi},\altaffilmark{2} and
Takeshi \textsc{Miyaji}\altaffilmark{2}}
\altaffiltext{1}{Graduate School of Science and Engineering, Kagoshima University,\\
                   1-21-35 K\^orimoto, Kagoshima, Kagoshima 890-0065}
\email{(KA) ando@milkyway.sci.kagoshima-u.ac.jp}
\altaffiltext{2}{Mizusawa VLBI Observatory, National Astronomical Observatory of Japan, \\
                 Osawa 2-21-1, Mitaka, Tokyo 181-8588}
\email{(TN) takumi.nagayama@nao.ac.jp}
\altaffiltext{3}{Institute of Astronomy, the University of Tokyo, Osawa 2-21-1, Mitaka, Tokyo 181-0015}
%\altaffiltext{3}{Mizusawa VERA Observatory, National Astronomical Observatory of Japan, \\
%                   2-12 Hoshi-ga-oka, Mizusawa-ku, Oshu, Iwate 023-0861}
%%% end:list of authors

%%% Please use the following style in case that sorting by 
%%% affilation is impossible. 
%
% \author{%
%   D-Firstname \textsc{D-Familyname}\altaffilmark{1}
%   E-Firstname \textsc{E-Familyname}\altaffilmark{1,2}
%   and
%   F-Firstname \textsc{F-Familyname}\altaffilmark{2}}
% \altaffiltext{1}{Address of Institute}
% \email{ddddd@xxx.xxx.xx.xx}
% \email{eeeee@xxx.xxx.xx.xx}
% \altaffiltext{2}{Address of Institute}

%% `\KeyWords{}' always has to be placed before `\maketitle'.
 %Do NOT move this preamble from here!

\maketitle

\begin{abstract}
We have performed the astrometry of H$_2$O masers 
in the Galactic star-forming region Onsala 2 North (ON2N)
with the VLBI Exploration of Radio Astrometry.
We obtained the trigonometric parallax of $0.261 \pm 0.009$ mas,
corresponding to the heliocentric distance of $3.83 \pm 0.13$ kpc.
ON2N is expected to be on the solar circle, because its radial velocity 
with respect to the Local Standard of Rest (LSR) is nearly zero.
Using present parallax and proper motions of the masers,
the Galactocentric distance of the Sun and 
the Galactic rotation velocity at the Sun are 
found to be $R_0 = 7.80\pm0.26$ kpc 
and $\Theta_0 = 213 \pm 5$ km s$^{-1}$, respectively.
The ratio of Galactic constants, namely the angular rotation velocity 
of the LSR
can be determined more precisely,
and is found to be 
$\Omega_0=\Theta_0/R_0 = 27.3\pm0.8$ km s$^{-1}$ kpc$^{-1}$,
which is consistent with the recent estimations 
but different from 25.9 km s$^{-1}$ kpc$^{-1}$ 
derived from the recommended values of $\Theta_0$ and $R_0$ 
by the International Astronomical Union (1985).
\end{abstract}

%%%%%%%%%%%%%%%%%%%%%%%%%%%%%%%%%%%%%%%%%%%%%%%%%%%%%%%%%%%%
%%%%%%%%%%%%%%%%%%%%%%%%%%%%%%%%%%%%%%%%%%%%%%%%%%%%%%%%%%%%
%%%%%%%%%%%%%%%%%%%%%%%%%%%%%%%%%%%%%%%%%%%%%%%%%%%%%%%%%%%%

\section{Introduction}

Very Long Baseline Interferometry (VLBI) astrometry is an important method
to measure the structure of the Milky Way Galaxy (MWG).
By measuring the accurate position of the source and its time variation,
the source distance and proper motion can be determined directly.
VLBI astrometry at 10 $\mu$as accuracy of 
the Galactic H$_2$O and CH$_3$OH maser sources
with the VLBI Exploration Radio Astrometry (VERA) and 
the Very Long Baseline Array (VLBA) can determine
accurate distances at kpc-scale with the errors less than 10\% 
(see for example \cite{hac06}; \cite{xu06}; \cite{hon07}).

The galactocentric distance of the Sun, $R_0$, and
the Galactic circular rotation velocity at the Sun, $\Theta_0$, 
are two fundamental parameters
to study the structure of the MWG 
and they are called here as the Galactic constants.
The rotation curve of the MWG and all kinematic distances of the sources
in the MWG are derived from these parameters.
Since 1985, the International Astronomical Union (IAU) has recommended
to give the values of $R_0=8.5$ kpc and $\Theta_0=220$ km s$^{-1}$.
However, recent studies report the values different from them 
(e.g. \cite{miyamoto98}; \cite{rei09a}).

Estimation of the Galactic constants is, however, affected
by several independent assumptions; the peculiar motion of the source,
systematic non-circular motions of 
both the source and the Local Standard of Rest (LSR) due to 
the spiral arm potential,
the non-axisymmetric potential of the MWG, 
the warping motion of the galactic disk,
and, furthermore, the motion of the Sun with respect to the LSR.
In this paper to simplify the situation we assume that the source moves with the perfect circular rotation on the disk.

\begin{figure}
  \begin{center}
    \FigureFile(80mm,80mm){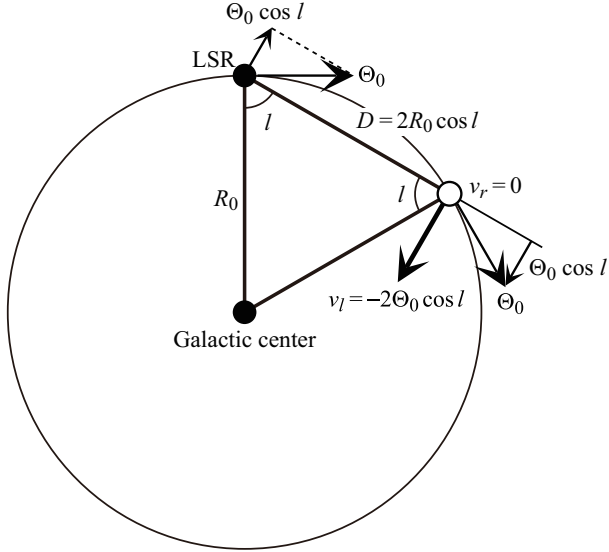}
  \end{center}
  \caption{The method to estimate $R_0$ and $\Theta_0$
            from the distance and the proper motion of 
            the source on the solar circle.}
  \label{fig:4}
\end{figure}

The circle with the raduis of the galactocentric distance $R_0$ of the Sun is
called as the solar circle.
All the sources on the circle perform the circular motion 
with the circular velocity $\Theta_0$ of the LSR,
provided that the non-circular motion of a source in the galactic disk
is negligible.
Due to the symmetric geometry, 
the radial velocity of a source on the circle 
is observed to be zero with respect to the LSR,
and the proper motion of the source depends only on 
the Galactic rotation velocity $\Theta_0$ of the LSR.
Therefore, we can derive $\Theta_0$ from the measured proper motion
of the source on the circle.
We can also derive $R_0$ from the heliocentric distance of the source, 
since the source, the Sun, and the Galactic center make 
an isosceles triangle (Figure \ref{fig:4}).
Thus, we can find directly the value of the angular velocity $\Omega_0$
of the LSR from $\Theta_0/R_0$. 
Traditionally, this value has been derived from the
Oort constants $A$ and $B$ on the basis of the kinematic analysis of stars
in the solar neighbourhood (see \cite{miyamoto98}).
The value of this ratio is a constraint 
to estimate one of the Galactic constants from the other.
Although the IAU gives recommended values of the Galactic constants,
at least one of them should be revised, 
if the ratio is inconsistent to the observed value.

Onsala 2 North (ON2N) is a massive star-forming region
located at the Galactic coordinates of $(l, b) = (\timeform{75.78D}, \timeform{0.34D})$.
Its radial velocity $v_{\mathrm{LSR}}$ with respect to the LSR is observed to be $0\pm1$ km s$^{-1}$ 
in the NH$_3$ and CS lines (\cite{olm99}; \cite{cod10}).
\citet{lek06} detected
the H$_2$O masers of ON2N at the radial velocity range from 
$-12$ to 9 km s$^{-1}$, with their peak flux densities of $10^2$--$10^3$ Jy.
These H$_2$O masers are associated with a 7 mm radio continuum source
and an NH$_3$ core as well, which are located at approximately 2" south 
from the ultracompact H\emissiontype{II} region, 
G75.78+0.34 (\cite{car97}; \cite{cod10}). 
Thus, ON2N is considered to be one of the sources on the solar circle.

We made astrometric observations of the H$_2$O masers in ON2N with VERA.
Based on these proper motions and trigonometric parallax measurement, 
we estimate
the Galactic constants $\Theta_0$, $R_0$, and $\Omega_0$.

In this paper we use the ``LSR velocity'' or $v_{\mathrm{LSR}}$ 
as the radial velocity with respect to the frame
moving toward $\alpha_{1900}=\timeform{18h}$, $\delta_{1900}=\timeform{+30D}$ 
with $-20$ km s$^{-1}$ namely, a provisional solar motion with respect to the LSR to be determined
after the traditional definition in radio astronomy since 1960s (see \cite{ker86}).

%%%%%%%%%%%%%%%%%%%%%%%%%%%%%%%%%%%%%%%%%%%%%%%%%%%%%%%%%%%%
%%%%%%%%%%%%%%%%%%%%%%%%%%%%%%%%%%%%%%%%%%%%%%%%%%%%%%%%%%%%
%%%%%%%%%%%%%%%%%%%%%%%%%%%%%%%%%%%%%%%%%%%%%%%%%%%%%%%%%%%%

\section{Observations and Data Reductions}

We observed H$_2$O masers at the rest frequency of 22.235080 GHz 
in the star-forming region 
ON2N with VERA at 14 epochs spanned about two years.
The day of year (DOY) of these epochs are 112, 206, 246, 305, 350 in 2006,
009, 051, 095, 132, 222, 277 in 2007, 005, 104, and 192 in 2008,
which are 53847, 53941, 53981, 54040, 54085, 54109, 54151, 54195, 54232,
54322, 54377, 54470, 54569, and 54657 in the Modified Julian Day (MJD), respectively.
The data of the first to third epochs were used to 
estimate the coodinates of H$_2$O masers.
We analyze the data of the rest 11 epochs in this study.
At each epoch, the H$_2$O masers in ON2N and 
a position reference continuum source, 
ICRF J201528.7+371059 (hereafter J2015+3710), were simultaneously observed 
in a dual-beam mode for about 10 hours.
The typical on-source integration time was 6 hours
for both ON2N and J2015+3710.
J2015+3710 is listed in the VLBA Calibrator Survey 2 (VCS2: \cite{fom03}).
J2015+3710 was detected with a peak flux density of 0.8--2 Jy in each epochs.
The separation angle between ON2N and J2015+3710 is \timeform{1.27D}.
The instrumental phase difference between the two beams was measured continuously
during the observations by injecting artifical noise sources into both beams
(\cite{hon08a}).
Left-hand circular polarization signals were sampled with 2-bit quantization,
and filtered with the VERA digital filter unit (\cite{igu05}).
The data were recorded onto magnetic tapes at a rate of 1024 Mbps,
providing a total bandwidth of 256 MHz, which consists of 16 IF channels with
a band width of 16 MHz each.
One IF channel was assigned to ON2N, and the other 15 IF
channels were assigned to J2015+3710, respectively.
Correlation processing was carried out on the Mitaka FX correlator.
The frequency and velocity spacings for ON2N
were 15.625 kHz and 0.21 km s$^{-1}$, respectively.

Data reduction was conducted using 
the NRAO Astronomical Image Processing System (AIPS).
An amplitude calibration was performed using 
the system noise temperatures during the observations.
For phase-referencing, a fringe fitting was
made using the AIPS task FRING on J2015+3710
with a typical integration time of 1 min 
and a time interval of 30 sec.
The solutions of the fringe phases, group delays, and delay rates
were applied to ON2N in order to calibrate the visibility data.
Phase and amplitude solutions obtained from 
self-calibration of J2015+3710 were also applied to ON2N.
Visibility phase errors caused by the Earth's atmosphere were calibrated
based on GPS measurements of the atmospheric zenith delay which occurs
due to tropospheric water vapor (\cite{hon08b}).
After the calibration, 
we made spectral-line image cubes 
each of which extent on the sky is $1024\times1024$ pixels with 0.05 mas
using the AIPS task IMAGR.
The typical size of the synthesized beam was $1.2\times0.9$ mas 
with position angle of \timeform{-50D}.
The rms noise for each image was approximately 0.1--1 Jy beam$^{-1}$.
The signal-to-noize ratio of 7 was adopted as the detection criterion.

%%%%%%%%%%%%%%%%%%%%%%%%%%%%%%%%%%%%%%%%%%%%%%%%%%%%%%%%%%%%
%%%%%%%%%%%%%%%%%%%%%%%%%%%%%%%%%%%%%%%%%%%%%%%%%%%%%%%%%%%%
%%%%%%%%%%%%%%%%%%%%%%%%%%%%%%%%%%%%%%%%%%%%%%%%%%%%%%%%%%%%

\section{Results and Discussion}

\subsection{Overall Properties of H$_2$O Masers in ON2N}

\begin{figure*}
  \begin{center}
    \FigureFile(160mm,160mm){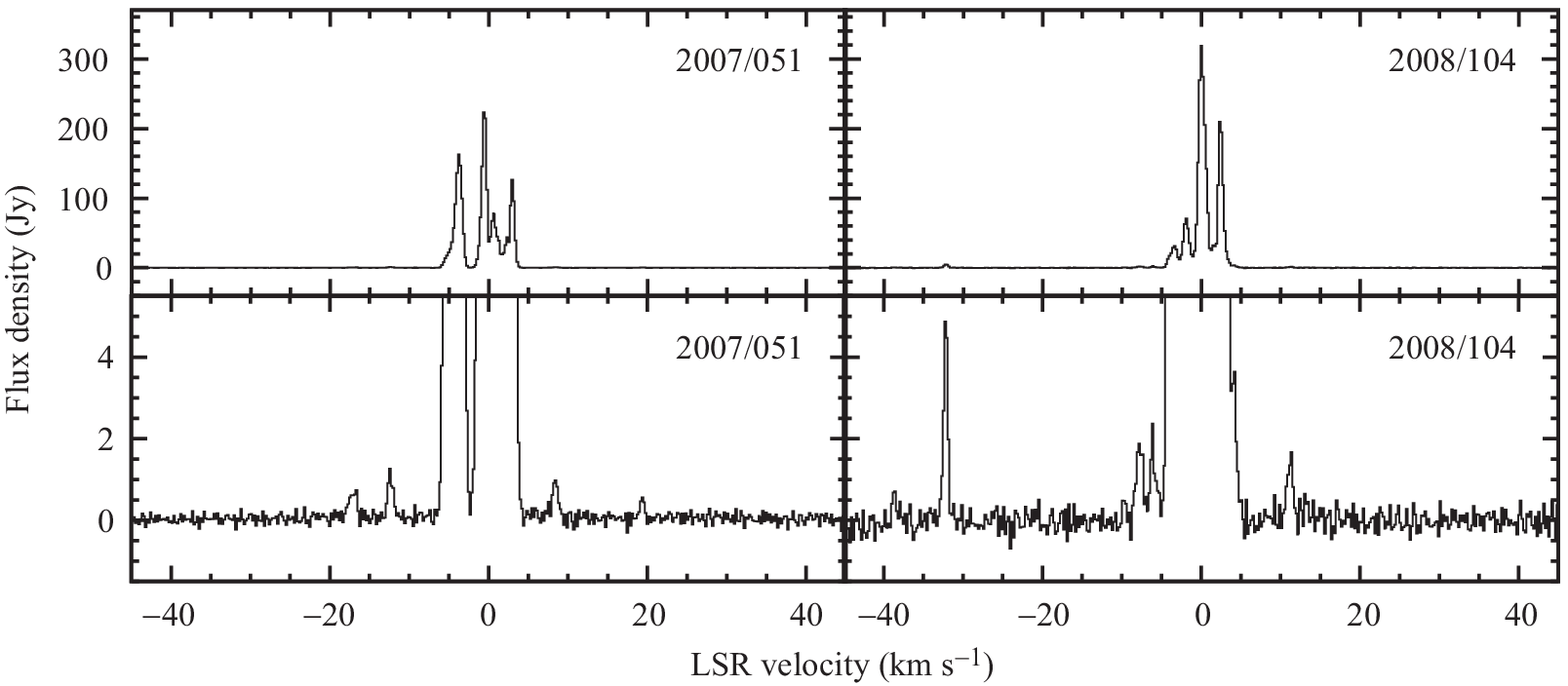}
  \end{center}
  \caption{A scalar-averaged cross-power spectra of H$_2$O masers in ON2N
           observed with the VERA Mizusawa-Iriki baseline at 2007/051 and 2008/104.}
  \label{fig:1}
\end{figure*}

Figure \ref{fig:1} shows the scalar-averaged cross-power spectra of 
H$_2$O masers in ON2N observed with the VERA Mizusawa-Iriki baseline on
2007/051 and 2008/104.
The intense emissions with the flux density of $\geq 100$ Jy
were detected at the LSR velocity range from 
$-5$ to 5 km s$^{-1}$.
The center of the velocity range of the intense emissions is close to 
the LSR velocity of the associated molecular cloud
at $v_{\rm LSR} = 0\pm1$ km s$^{-1}$ 
of the NH$_3$ and CS lines (\cite{olm99}; \cite{cod10}).
The majority of $v_{\mathrm{LSR}}$ of H$_2$O masers is found in the range
$-12 \leq v_{\rm LSR} \leq 9$ km s$^{-1}$ seen in the previous
monitoring observations in 1995--2004 (\cite{lek06}).
We also detected the blueshifted components 
at $v_{\rm LSR} = -33$ and $-17$ km s$^{-1}$ and the 
redshifted component at 19 km s$^{-1}$,
which were not detected in the 
preveous observations (e.g. \cite{lek06}). 

\begin{figure*}
  \begin{center}
    \FigureFile(160mm,160mm){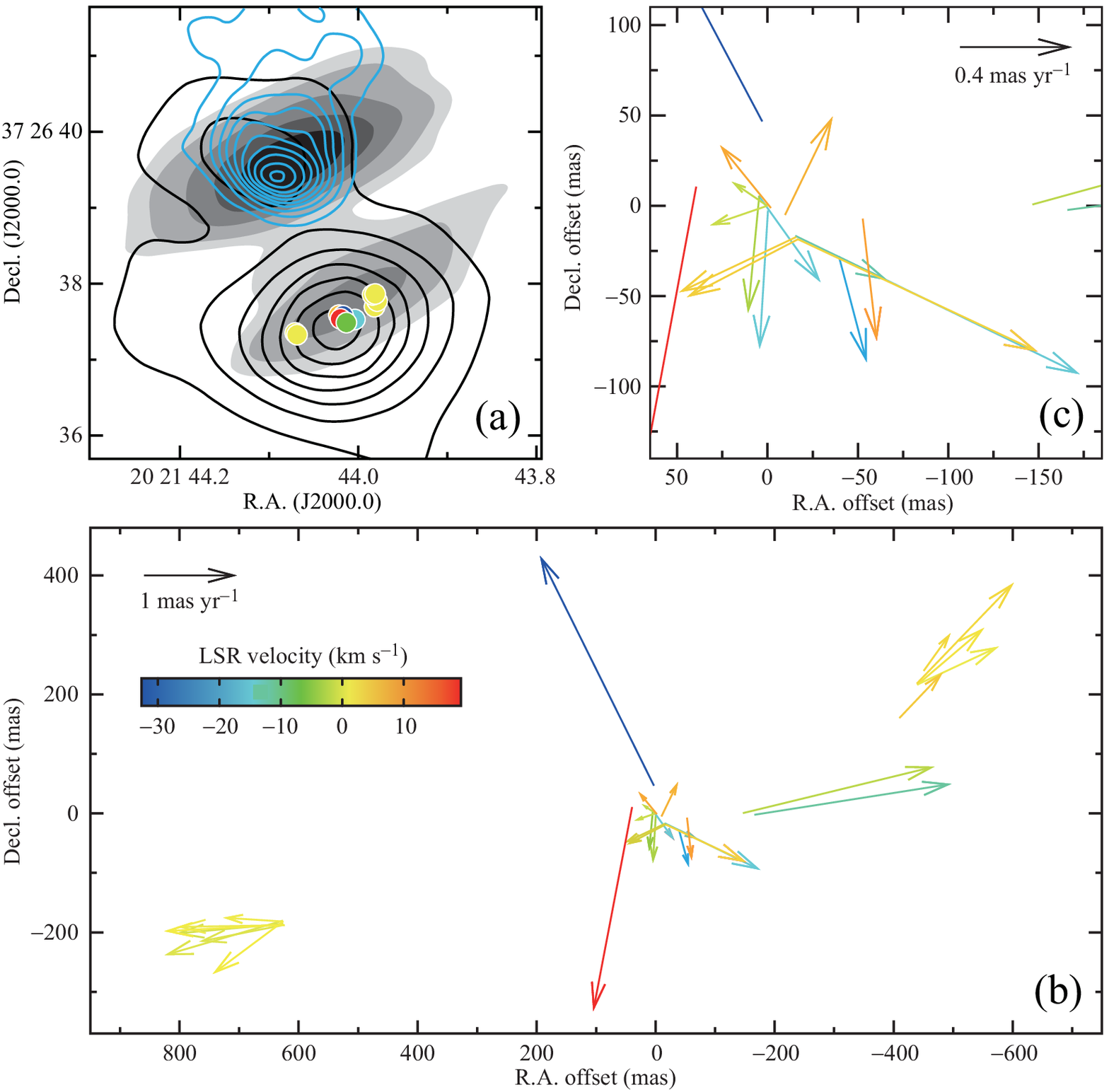}
  \end{center}
  \caption{(a): Distributions of H$_2$O masers (color filled circles) 
            superimposed on the 6 cm radio continuum emission (cyan contour; \cite{woo89}),
            the 7 mm radio continuum emission (gray contour; \cite{car97}),
            and the NH$_3$ (3,3) emission (black contour; \cite{cod10}).
            (b): Internal motion vectors of H$_2$O masers.
            The spot color shows the LSR velocity. The arrow at the top left corner
            shows a internal motion of 1 mas yr$^{-1}$, corresponding to 18.2 km s$^{-1}$
            at a distance of 3.83 kpc.
            (c): Close-up to the central part of (b).}
  \label{fig:2}
\end{figure*}

Thirty H$_2$O maser spots were detected over the half of year and at more than three epochs.
They were in the LSR velocity range from $-33$ to 19 km s$^{-1}$, 
and distrubuted with an area of 1.0"$\times$0.4".
Figure \ref{fig:2} shows the distribution of internal motion
of the maser spots in ON2N.
The reference position of the map is set at the position of a maser spot at
$v_{\rm LSR} = 0.1$ km s$^{-1}$, which is estimated to be
$(\alpha, \delta)_{\rm J2000.0} = 
(\timeform{20h21m44.01225s}, \timeform{37D36'37.4844"})$.
Although the H$_2$O masers are located at 2" south from the peak of the 6 cm
radio continuum emission (\cite{woo89}),
they are spatially coincident with the peaks of the 7 mm radio 
continuum emission and the NH$_3$ (3,3) emission (\cite{car97}; \cite{cod10}).

\subsection{Parallax and Proper Motion}

The absolute motion of the respective H$_2$O maser spot, 
i.e. its motion with respective to 
the position reference source J2015+3710,
is given by the sum of the proper motion and the annual parallactic motion.
In order to separate these two motions
we performed monitoring observations of the H$_2$O masers of ON2N 
for about two years.

\begin{figure*}
  \begin{center}
    \FigureFile(160mm,80mm){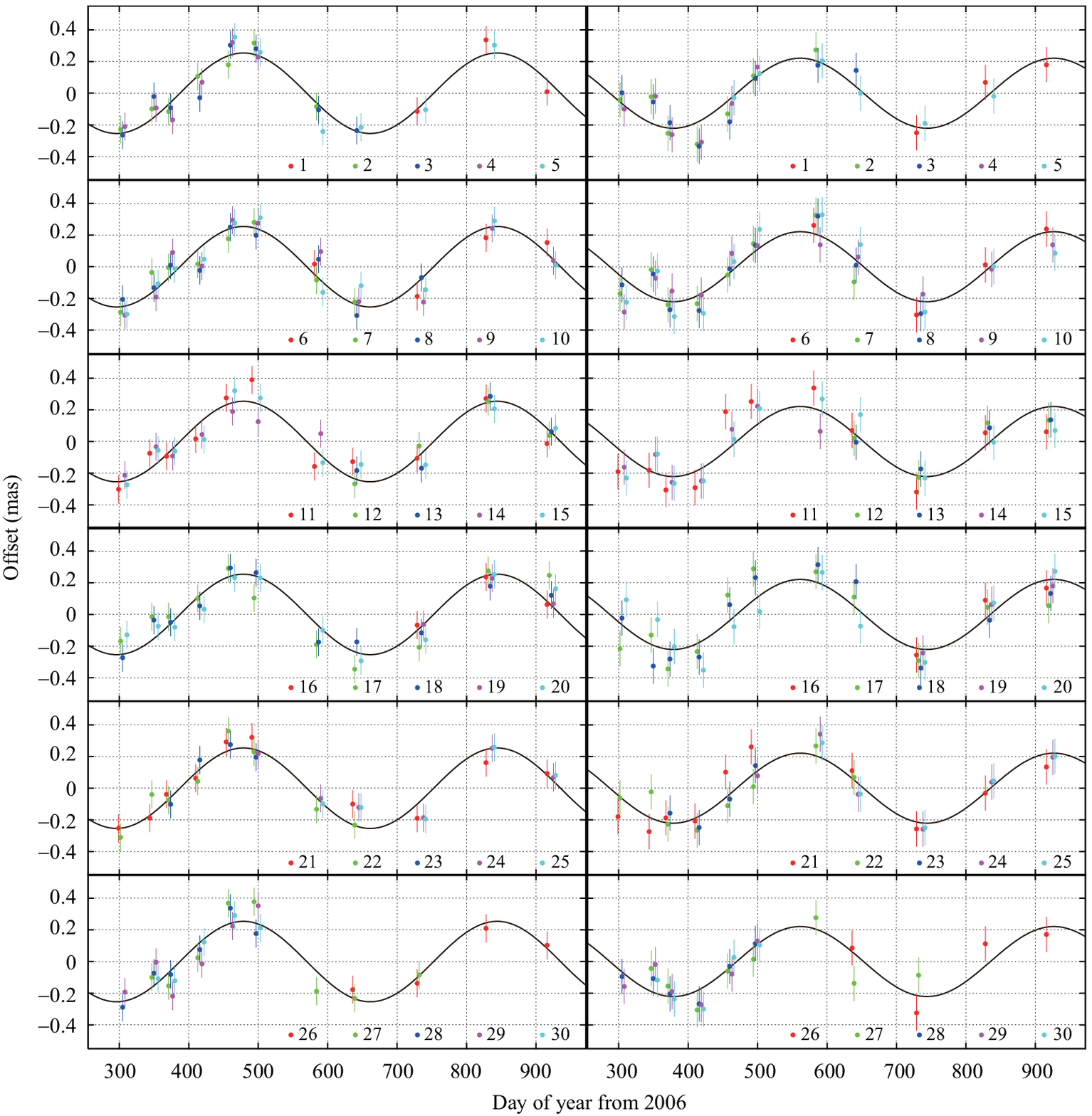}
  \end{center}
  \caption{Parallax of the H$_2$O masers in ON2N.
  The data for the different maser positions are slightly shifted
  in time for clarity.
  Individual proper motions and position offsets are removed.
  The left and right panels show the annual parallax in right ascension and declination, respectively.
  The numbers in each panel show the ID number of the spot listed in Table \ref{tab:1}.}
  \label{fig:3}
\end{figure*}

We made a combined parallax fit, 
which means a fitting of the positions of 30 H$_2$O maser spots to
a common parallax, but different proper motion and 
position offset for each spot.
Figure \ref{fig:3} and Table \ref{tab:1} show 
the results of the combined parallax fit.
As can be clearly seen in Figure \ref{fig:3},
the observed points demonstrate a sinusoidal modulation with a period of 1 year
caused by the annual parallax.
For this fitting,
we assigned independent ``error floors'' in quadrature 
with the formal position fitting uncertainties.
Trial combined fits were conducted and a separate reduced $\chi^2$
(per degree of freedom) statistics was applied 
for the right ascension and declination residuals.
The error floors of 0.088 mas and 0.111 mas in the right ascension and the declination respectively,
were then adjusted iteratively so as to 
achive a reduced $\chi^2$ per dgree of freedom near unity 
in each coordinate.
Combining all of the fittings, we obtained the trigonometric parallax
of the H$_2$O maser spots $0.261\pm0.009$ mas. 
The parallax gives the heliocentric distance of ON2N $3.83\pm0.13$ kpc.

In Table \ref{tab:1},
we also show the estimated parallaxes using individual fitting for each maser spot.
We made this individual fitting only for 14 maser spots 
which were detected over one year.
The obtained parallaxes from the individual fitting are
consistent with each other 
and with the result of the combined fit.
This means that the parallax obtained by the combined fitting is reliable.

The systemic motion of the source can be estimated as 
an averaged motion of all maser spots,
provided that internal motion is random or symmetric.
We believe that this may be reasonable for ON2N because of the following two reasons.
The averaged radial velocity of all maser spots is $-1.2$ km s$^{-1}$  
which is close to the systemic radial velocity derived from the associated molecular cloud.
This suggests that the maser spots move rather symmetrically.
Figure \ref{fig:2} shows the residual proper motion vectors, 
which are differences between the individual proper motions and the average. 
We did not find any strong asymmetric motion.
Therefore, 
we find that the reasonable absolute proper motions 
are not biased by the internal motions of ON2N.
Thus, the systemic proper motion of ON2N is estimated to be
$(\mu_{\alpha} \cos\delta, \mu_{\delta}) = 
(-2.79 \pm 0.13, -4.66 \pm 0.17)$ mas yr$^{-1}$
as the average motion of all 30 maser spots.

Using the Galactic coordinates of ON2N 
$(l, b) = (\timeform{75.78D}, \timeform{-0.34D})$,
the proper motion components in the galactic coordinates are calculated to be
$(\mu_{l} \cos b, \mu_{b}) = (-5.42\pm0.16, -0.36 \pm 0.14)$ mas yr$^{-1}$
corresponding to the linear velocity of 
$(v_{l}, v_{b}) = (-98.4\pm2.9, -6.6 \pm 2.6)$ km s$^{-1}$ 
at the distance of 3.83 kpc.
We note that these values are still affected by the solar motion, 
because the observed proper motion is not relative to the LSR but to the Sun.

To convert this observed velocity to that with respect to the LSR,
we have to fix the solar motion relative to the LSR.
As mentioned in \S 1 we use the solar motion in the traditional definition of
$(U_\odot, V_\odot, W_\odot)=(+10.3, +15.3, +7.7)$ km s$^{-1}$ 
after \citet{rei09a}
\footnote {This velocity is consistent with the provisional solar motion we used,
but the 3 dimensional linear velocity 
$(U_\odot, V_\odot, W_\odot)=(+10.0, +15.4, +7.8)$ km s$^{-1}$
shown in \citet{ker86} is inconsistent to the definition by themselves.}.
We note that this set of values is very close to the one found by 
\citet{miyamoto98} (see Table 5) on the basis of the stellar motion
analysis of the HIPPARCOS proper motions in the solar neighbourhood.
Using this traditional solar motion, the observed proper motion is converted to
$(\mu_{l} \cos b, \mu_{b}) = (-5.76 \pm0.16, 0.02 \pm0.14)$ mas yr$^{-1}$ 
and the corresponding linear velocity is
$(v_{l}, v_{b}) = (-104.6 \pm2.9, 1.1 \pm 2.6)$ km s$^{-1}$
relative to the LSR.

\begin{table*}
  \caption{The obtained values of parallax $\pi$ and proper motions $\mu_{\alpha \cos\delta}$ and $\mu_{\delta}$
           for H$_2$O maser features in ON2N.}
  \label{tab:1}
  \begin{center}
    \begin{tabular}{rrrrcrrr}
      \hline
                                            &
      \multicolumn{1}{c}{$v_{\rm LSR}$}   &
      \multicolumn{1}{c}{$\Delta \alpha \cos\delta$}             &
      \multicolumn{1}{c}{$\Delta \delta$}             &
                                            &
     \multicolumn{1}{c}{$\pi$}           &
      \multicolumn{1}{c}{$\mu_{\alpha \cos \delta}$}         &
      \multicolumn{1}{c}{$\mu_{\delta}$}         \\
      \multicolumn{1}{c}{ID}              &
      \multicolumn{1}{c}{(km s$^{-1}$)}   &
      \multicolumn{1}{c}{(mas)}           &
      \multicolumn{1}{c}{(mas)}           &
      \multicolumn{1}{c}{Epochs}          &
      \multicolumn{1}{c}{(mas)}           &
      \multicolumn{1}{c}{(mas yr$^{-1}$)} &
      \multicolumn{1}{c}{(mas yr$^{-1}$)} \\ 
     \hline
1&   $-32.8$ &$   3.0$&$  46.7$&\texttt{........IJK}&-----          &$-1.53\pm0.24$&$-2.12\pm0.31$\\
2&   $-17.6$ &$ -39.5$&$ -28.1$&\texttt{ABCDEFG....}&-----          &$-2.89\pm0.14$&$-5.04\pm0.17$\\
3&   $-16.5$ &$ -39.0$&$ -28.2$&\texttt{ABCDEFGH...}&0.252$\pm$0.043&$-3.67\pm0.10$&$-5.09\pm0.13$\\
4&   $-12.1$ &$ -15.3$&$ -16.7$&\texttt{ABCDEF.....}&-----          &$-3.12\pm0.21$&$-4.82\pm0.26$\\
5&   $-10.4$ &$-165.9$&$  -2.4$&\texttt{....EFGHIJ.}&0.258$\pm$0.024&$-4.96\pm0.10$&$-4.32\pm0.13$\\
6&   $ -4.8$ &$  -0.2$&$  -2.6$&\texttt{......G.IJK}&-----          &$-2.76\pm0.12$&$-5.16\pm0.16$\\
7&   $ -4.8$ &$   4.8$&$   5.5$&\texttt{ABCDEFGH...}&0.236$\pm$0.036&$-3.01\pm0.10$&$-4.97\pm0.13$\\
8&   $ -3.7$ &$   4.6$&$   5.4$&\texttt{ABCDEFGHI..}&0.257$\pm$0.037&$-2.75\pm0.08$&$-5.08\pm0.10$\\
9&   $ -1.8$ &$-146.6$&$   0.6$&\texttt{ABCDEFGHIJK}&0.229$\pm$0.038&$-4.90\pm0.04$&$-4.15\pm0.06$\\
10&  $ -1.2$ &$   0.2$&$   0.0$&\texttt{ABCDEFGHIJK}&0.276$\pm$0.031&$-2.67\pm0.05$&$-4.58\pm0.06$\\
11&  $  0.1$ &$   0.0$&$   0.0$&\texttt{ABCDEFGHIJK}&0.287$\pm$0.039&$-2.59\pm0.05$&$-4.73\pm0.06$\\
12&  $  0.5$ &$ 626.1$&$-185.4$&\texttt{.......HIJK}&-----          &$-1.62\pm0.15$&$-4.77\pm0.19$\\
13&  $  0.7$ &$ 625.2$&$-187.6$&\texttt{.......HIJK}&-----          &$-1.88\pm0.15$&$-4.84\pm0.19$\\
14&  $  0.7$ &$ 626.2$&$-180.2$&\texttt{ABCDEFG....}&-----          &$-1.50\pm0.14$&$-5.04\pm0.17$\\
15&  $  1.4$ &$-439.8$&$ 216.2$&\texttt{ABCDEFGHIJK}&0.252$\pm$0.031&$-3.51\pm0.05$&$-4.05\pm0.06$\\
16&  $  1.4$ &$ 623.2$&$-187.9$&\texttt{........IJK}&-----          &$-1.62\pm0.24$&$-4.68\pm0.30$\\
17&  $  1.6$ &$ 627.7$&$-181.7$&\texttt{ABCDEFGHIJK}&0.272$\pm$0.048&$-2.17\pm0.05$&$-4.62\pm0.06$\\
18&  $  1.8$ &$-439.5$&$ 216.4$&\texttt{ABCDEFGHIJK}&0.275$\pm$0.031&$-3.67\pm0.05$&$-4.25\pm0.06$\\
19&  $  1.8$ &$ 625.1$&$-187.3$&\texttt{........IJK}&-----          &$-1.49\pm0.24$&$-4.73\pm0.31$\\
20&  $  1.8$ &$ 625.8$&$-181.2$&\texttt{ABCDEFGHIJK}&0.250$\pm$0.034&$-2.03\pm0.05$&$-5.23\pm0.06$\\
21&  $  2.0$ &$-450.5$&$ 239.4$&\texttt{ABCDEFG.IJK}&0.264$\pm$0.038&$-3.07\pm0.05$&$-4.27\pm0.06$\\
22&  $  2.0$ &$-439.1$&$ 218.2$&\texttt{ABCDEFGH...}&0.270$\pm$0.043&$-3.85\pm0.10$&$-3.56\pm0.13$\\
23&  $  2.2$ &$ -17.1$&$ -18.7$&\texttt{..CDEF.....}&-----          &$-3.66\pm0.35$&$-5.07\pm0.44$\\
24&  $  2.4$ &$ -15.8$&$ -17.4$&\texttt{.....FGHIJK}&-----          &$-2.37\pm0.09$&$-4.86\pm0.11$\\
25&  $  2.6$ &$ -17.1$&$ -18.4$&\texttt{......GHIJK}&-----          &$-2.39\pm0.11$&$-4.87\pm0.15$\\
26&  $  3.3$ &$-409.6$&$ 160.2$&\texttt{.......HIJK}&-----          &$-3.25\pm0.15$&$-4.18\pm0.19$\\
27&  $  5.8$ &$  -9.7$&$  -5.1$&\texttt{.BCDEFGHI..}&0.292$\pm$0.050&$-2.96\pm0.09$&$-4.31\pm0.11$\\
28&  $  8.3$ &$  -1.9$&$  -1.2$&\texttt{ABCDEF.....}&-----          &$-2.61\pm0.20$&$-4.44\pm0.26$\\
29&  $  8.7$ &$ -52.7$&$  -7.2$&\texttt{ABCDEF.....}&-----          &$-2.84\pm0.21$&$-5.10\pm0.25$\\
30&  $ 19.3$ &$  39.5$&$  10.6$&\texttt{.BCDEF.....}&-----          &$-2.36\pm0.27$&$-6.91\pm0.34$\\
      \hline
      \multicolumn{5}{c}{Combined fit}  &
      $0.261 \pm 0.009$                  &
                                          &
                                         \\
      \multicolumn{5}{c}{Average}       &
                                          &
      $-2.79 \pm 0.13$                   &
      $-4.66 \pm 0.17$                   \\
  \hline
\multicolumn{8}{@{}l@{}} {\hbox to 0pt{\parbox{150mm}{\footnotesize
Column (3), (4): Right ascension and declination offsets 
       relative to the positon of the maser spot at 
       $v_{\rm LSR} = 0.1$ km s$^{-1}$, 
       and $(\alpha, \delta)_{\rm J2000.0} = 
       (\timeform{20h21m44.01225s}, \timeform{37D36'37.4844"})$.\\
Column (5): Each alphabetical letter represents the epoch with maser detection. 
A, B, C, ..., and K mean the 11 epochs from 53847, 53941, ..., and 54657 in MJD, respectively.
A dot represents the epoch without the detection. \\
Column (6): Parallax estimated from the individual fitting.\\
Column (7), (8): Motions on the sky in the directions 
along the right ascension and declination.\\
}\hss}}
    \end{tabular}
  \end{center}
\end{table*}

\subsection{Derivation of the Galactic Constants}

Based on the observed LSR velocity of ON2N 
we believe that the source is located at or close to the solar circle.
The small proper motion along the galactic latitude, 
$v_b=1.1\pm2.6$ km s$^{-1}$, shown in \S 3 supports 
that ON2N rotates circularly around the Galactic center.

For an object on the solar circle, the galactocentric distance
of the Sun or that of the source, $R_0$, is estimated from 
the heliocentric distance of the source as
\begin{equation}
R_0 = \frac{D}{2 \cos l},
\label{equ:1} 
\end{equation}
where $l$ is the galactic longitude of the source (see Figure \ref{fig:4}).
Our estimation of the heliocentric distance of ON2N yields $D=3.83\pm0.13$ kpc.
It gives $R_0=7.80\pm0.26$ kpc, if ON2N is exactly located on the solar circle.
This value is close to the previous estimations;
$R_0$ is estimated to be $8.0\pm0.5$ kpc
from a combination of many methods reviewd by \citet{rei93},
$7.9^{+0.8}_{-0.7}$ kpc from a parallax measurement 
of H$_2$O masers in Sgr B2 with VLBA (\cite{rei09b}),
and $8.28\pm0.44$ kpc using the orbits of stars around Sgr A* 
from VLT and Keck data (\cite{gil09}).

For a source on the solar circle with purely circular rotation,
its proper motion velocity along the galactic longitude, $v_l$, gives
the galactic rotation velocity of the LSR or that of the object as
\begin{equation}
\Theta_0 = - \frac{v_l}{2 \cos l},
\label{equ:2} 
\end{equation}
where $l$ is the galactic longitude of the source.
Our measured value of $v_l=-104.6\pm2.6$ km s$^{-1}$ gives
$\Theta_0=213\pm5$ km s$^{-1}$.
This value is smaller than that estimated by \citet{rei09a},
$\Theta_0=254\pm16$ km s$^{-1}$ but close to the IAU recommended
value of $\Theta_0=220$ km s$^{-1}$.

Our derivations of the Galactic constants are strongly
dependent on the assumption of the location of ON2N in the MWG.
However, we found that the ratio of the Galactic constants, $\Theta_0 / R_0$,
which is the anglular velocity of the LSR, $\Omega_0$, can be estimated
less dependently on the assumption.
The ratio $\Theta_0 / R_0$ can be estimated, 
even if ON2N is not exactly located on the solar circle, but near there.

For a source on pure circular rotation at any position in the galactic disk,
its radial and tangential velocities with respect to the LSR
can be written as
\begin{eqnarray}
v_r &=& \left( \frac{\Theta}{R} - \frac{\Theta_0}{R_0} \right) R_0 \sin l,
\label{equ:3} \\
v_l &=& \left( \frac{\Theta}{R} - \frac{\Theta_0}{R_0} \right) R_0 \cos l - \frac{\Theta}{R}D,
\label{equ:4}
\end{eqnarray}
where $R$ is the actual galactocentric distance of the source, 
and $D$ is the heliocentric distance of the source,
and $\Theta$ is the galactic rotation velocity of the source.
Equations (\ref{equ:3}) and (\ref{equ:4}) yield
\begin{eqnarray}
\nonumber
\frac{\Theta_0}{R_0} &=& -\frac{v_l}{D} + v_r \left(\frac{1}{D \tan l} - \frac{1}{R_0 \sin l}\right) \\
&=& -a_0 \mu_l + v_r \left(\frac{1}{D \tan l} - \frac{1}{R_0 \sin l}\right),
\label{equ:7} 
\end{eqnarray}
where $a_0$ is a constant to convert the unit from an angular velocity to a linear velocity and
$a_0=4.74$ km s$^{-1}$ mas$^{-1}$ yr kpc$^{-1}$.
For a source near the solar circle, its $v_r$ 
is nearly zero.
In this case, equation (\ref{equ:7}) yields 
\begin{equation}
\frac{\Theta_0}{R_0} \simeq -a_0 \mu_l,
\end{equation}
which is free from $R_0$.
Actually we found that $\Theta_0$/$R_0$ is a nearly constant at $6 \leq R_0 \leq 10$ kpc
and the ratio is obtained as $\Theta_0 / R_0 = 27.3 \pm 0.8$ km s$^{-1}$ kpc$^{-1}$
using the $D = 3.83\pm0.13$ kpc, $\mu_l = -5.76 \pm 0.16$ mas yr$^{-1}$, 
and $v_r = 0 \pm 1$ km s$^{-1}$.
This value is close to the value of 
$\Theta_0 / R_0 = 28.7 \pm 1.3$ km s$^{-1}$ kpc$^{-1}$
obtained from a tangent point source, ON1 (\cite{nag11}),
and the value of
$\Theta_0 / R_0 = 28.6 \pm 0.2$ km s$^{-1}$ kpc$^{-1}$
obtained from the proper motion measurement of Sgr A* (\cite{rei04}),
which is revised using the traditional definition of the solar motion by us.
However, this value is inconsistent to that derived from the IAU recommended values
220 km s$^{-1}$ / 8.5 kpc=25.9 km s$^{-1}$ kpc$^{-1}$.
This estimation gives another constraint 
on the Galactic constants which is independent of the Oort constants 
derived from stellar motion near the Sun.

%%%%%%%%%%%%%%%%%%%%%%%%%%%%%%%%%%%%%%%%%%%%%%%%%%%%%%%%%%%%
%%%%%%%%%%%%%%%%%%%%%%%%%%%%%%%%%%%%%%%%%%%%%%%%%%%%%%%%%%%%
%%%%%%%%%%%%%%%%%%%%%%%%%%%%%%%%%%%%%%%%%%%%%%%%%%%%%%%%%%%%

\bigskip
We thank the referee Dr. Masanori Miyamoto for his invaluable comments and suggestions.
We also thank to the staff members of all the VERA stations
for their assistances in the observations.

%%%%%%%%%%%%%%%%%%%%%%%%%%%%%%%%%%%%%%%%%%%%%%%%%%%%%%%%%%%%
%%%%%%%%%%%%%%%%%%%%%%%%%%%%%%%%%%%%%%%%%%%%%%%%%%%%%%%%%%%%
%%%%%%%%%%%%%%%%%%%%%%%%%%%%%%%%%%%%%%%%%%%%%%%%%%%%%%%%%%%%

%%%
% See the manual for the detail.
%%%

\end{document}